\documentclass[12pt]{article}

\baselineskip=\normalbaselineskip

\usepackage{mathrsfs,amsbsy,amssymb,latexsym,amsfonts,amsmath,amsthm}
\usepackage{graphicx,color}

\makeatletter
\catcode`\@=11
\@addtoreset{equation}{section}

\def\@seccntformat#1{\csname the#1\endcsname.~~}
\makeatother

\newcommand{\bbC}{{\mathbb{C}}}
\newcommand{\bbR}{{\mathbb{R}}}

\begin{document}

\begin{titlepage}
\renewcommand{\thefootnote}{\fnsymbol{footnote}}
\begin{flushright}
KUNS-2666
\end{flushright}
\vspace*{1.0cm}

\begin{center}
{\Large \bf 
Parallel tempering algorithm \\
for integration over Lefschetz thimbles
}
\vspace{1.0cm}

\centerline{
{Masafumi Fukuma}%
\footnote{E-mail address: 
fukuma@gauge.scphys.kyoto-u.ac.jp} 
and
{Naoya Umeda}%
\footnote{E-mail address: 
n\_umeda@gauge.scphys.kyoto-u.ac.jp}%
}

\vskip 0.8cm
{\it Department of Physics, Kyoto University, Kyoto 606-8502, Japan}
\vskip 1.2cm 

\end{center}

\begin{abstract}

The algorithm based on integration over Lefschetz thimbles 
is a promising method 
to resolve the sign problem for complex actions. 
However, this algorithm often meets a difficulty 
in actual Monte Carlo calculations 
because the configuration space is not easily explored 
due to the infinitely high potential barriers between different thimbles. 
In this paper, 
we propose to use the flow time of the antiholomorphic gradient flow 
as an auxiliary variable for the highly multimodal distribution. 
To illustrate this, 
we implement the parallel tempering method 
by taking the flow time as a tempering parameter.  
In this algorithm, 
we can take the maximum flow time 
to be sufficiently large such that the sign problem disappears there, 
and two separate modes are connected 
through configurations at small flow times. 
To exemplify that this algorithm does work, 
we investigate the $(0+1)$-dimensional massive Thirring model at finite density 
and show that our algorithm correctly reproduces the analytic results 
for large flow times such as $T=2$. 

\end{abstract}
\end{titlepage}

\pagestyle{empty}
\pagestyle{plain}

\tableofcontents
\setcounter{footnote}{0}

\section{Introduction}
\label{Introduction}

There are many cases where one needs to deal with complex actions. 
One important example for high energy/nuclear physics is 
quantum chromodynamics (QCD) at finite density 
(see Ref. \cite{Aarts:2015tyj} for a review on the recent developments). 
However, since a complex action does not give  
a real and positive Boltzmann weight, 
one cannot directly resort to the traditional Markov chain Monte Carlo methods 
to estimate correlation functions. 
The so-called reweighting algorithm 
(which absorbs the phase of the weight into observables)
is highly ineffective 
when the imaginary part of the action becomes very large 
(such as in the thermodynamic limit), 
because one needs to take a sample from a configuration space  
where the weights of nearby configurations have almost the same amplitudes 
but very different phases. 
The difficulty of numerical evaluation in such a situation 
is termed the {\em sign problem}.

There have been many proposals to circumvent the sign problem. 
One of the approaches that are currently under intense study 
is the use of integration over Lefschetz thimbles 
\cite{Cristoforetti:2012su}
(see also 
Refs. \cite{Cristoforetti:2013wha,Mukherjee:2013aga,
Fujii:2013sra,Cristoforetti:2014gsa,Alexandru:2015xva,
Kanazawa:2014qma,Tanizaki:2015rda,Fujii:2015bua}), 
which we will call the Lefschetz thimble algorithm hereafter. 
There, the original real-valued variable (say, $x=(x^i)\in\bbR^N$)
is complexified according to the {\em antiholomorphic gradient flow} 
(sometimes called the {\em upward flow} in the literature)  
$\dot{z}^i=\left[\partial_i S(z)\right]^\ast$. 
In the original Lefschetz thimble algorithm, 
as will be reviewed in Sect.\ \ref{subsec:Lefschetz}, 
the flow time is taken to infinity 
to map the original configuration space 
to a union of Lefschetz thimbles. 
Since the imaginary part of the action is constant on each thimble, 
the sign problem disappears 
if the path integral is made over the Lefschetz thimbles. 
However, since two different thimbles are separated 
by an infinitely high potential barrier, 
one needs to invent some machinery 
to incorporate contributions from all relevant thimbles.

Recently, Alexandru et al.\ \cite{Alexandru:2015sua} 
was a very interesting proposal 
to consider configurations on a manifold 
that is obtained from the original configuration manifold 
by a finite amount of flow time. 
This algorithm was a great success in various models 
\cite{Alexandru:2015sua,Alexandru:2016ejd}, 
but reducing the amount of flow time 
may also reduce the effectiveness against the sign problem, 
and one does not know a priori 
whether the chosen flow time avoids 
both the sign problem and the multimodal problem simultaneously.

In this paper, as a versatile tool 
for Monte Carlo calculations of models with complex actions,  
we propose a Lefschetz thimble algorithm 
where the flow time is used as an {\em auxiliary variable}  
for the highly multimodal distribution.  
There can be various methods to realize this idea, 
and in this paper we implement the parallel tempering method 
because of its simplicity,  
by taking the flow time as a tempering parameter. 
There, 
we consider a set of manifolds corresponding to various flow times. 
Two separate modes at large flow times 
(where the sign problem no longer exists) 
are then connected by passing through configurations at small flow times 
(where the original sign problem exists 
but the multimodality is expected to be mild). 
Since the sample average is taken only with respect to the largest flow time, 
we need not worry about the sign problem at small flow times, 
although we take into account configurations there.

This paper is organized as follows. 
In Sect. \ref{sec:Algorithm} 
we first review the basics of the Lefschetz thimble algorithm 
based on Refs. \cite{Alexandru:2015xva,Alexandru:2015sua}, 
and then implement the parallel tempering method in the algorithm 
by taking the flow time as a tempering parameter. 
In Sect. \ref{sec:Example} 
we investigate the $(0+1)$-dimensional massive Thirring model at finite density, 
and show that our algorithm correctly reproduces the analytic results 
for large flow times such as $T=2$. 
Section \ref{sec:Conclusion} is devoted 
to the conclusion and outlook for future work.

\section{Algorithm}
\label{sec:Algorithm}

\subsection{Integration over Lefschetz thimbles (review)}
\label{subsec:Lefschetz}

We consider a real $N$-dimensional dynamical variable, $x=(x^i)\in\bbR^N$, 
with action $S(x)$ that may take complex values for real-valued $x$. 
Our main concern is to evaluate the expectation values 
of functions of $x$: 
\begin{align}
 \langle \mathcal{O}(x) \rangle 
 = \frac{\int_{\bbR^N} d x\, e^{-S(x)}\,\mathcal{O}(x)}
 {\int_{\bbR^N} d x\, e^{-S(x)}}.
\label{eq:vev1}
\end{align}
We assume 
that $|e^{-S(x)}|$ decreases rapidly enough in the limit $x\to \pm\infty$,  
and that $e^{-S(z)}$ and $\mathcal{O}(z)$ are entire functions 
when regarded as functions of $z=(z^i)\in \bbC^N$. 
The integration region can then be changed 
to any other region $\Sigma$ in $\bbC^N$ 
as long as it is obtained as a continuous deformation of 
the original region with the boundary fixed at infinity. 
We consider as such an integration region 
the submanifold that is obtained 
from the following antiholomorphic gradient flow $z(t;x)$ 
with a flow time $t$:
\begin{align}
 \frac{d z^i}{d t} &= \Bigl[\frac{\partial S(z)}{\partial z^i}\Bigr]^\ast,
 \quad
 z^i|_{t=0} = x^i. 
\end{align}
In fact, the flow defines a map from the original integration region 
$\Sigma_0\equiv\bbR^N$ 
to a real $N$-dimensional submanifold $\Sigma_t$ in $\bbC^N$: 
\begin{align}
 z_t\,:~\Sigma_0 \ni x ~\mapsto~ z_t(x)\equiv z(t;x) \in \Sigma_t .
\end{align}
We thus see that \eqref{eq:vev1} can be rewritten as
\begin{align}
 \langle \mathcal{O}(x) \rangle 
 = \frac{\int_{\Sigma_t} d  z\, e^{-S(z)}\,\mathcal{O}(z)}
 {\int_{\Sigma_t} d z\, e^{-S(z)}}, 
\label{eq:LT1}
\end{align}
which can be further rewritten as a reweighted integral over $\bbR^N$ 
\cite{Alexandru:2015xva} as%
\footnote{ 
One may also take as an integration region 
the tangent space to the critical point 
of the dominant thimble \cite{Alexandru:2015xva} 
if the integral can be well approximated 
by the integration around the critical point. 
} 
\begin{align}
 \langle \mathcal{O}(x) \rangle 
 &= \frac{\int_{\bbR^N} d x\, 
 \det\!J_t(x)\,e^{-S(z_t(x))}\,\mathcal{O}\bigl(z_t(x)\bigr)}
 {\int_{\bbR^N} d x\, \det\!J_t(x)\,e^{-S(z_t(x))}} 
\nonumber\\
 &=\frac{
 \Bigl\langle e^{i \bigl[{\rm arg}\det\!J_t(x)-S_I(z_t(x))\bigr]} 
 \mathcal{O}\bigl(z_t(x)\bigr)
  \Bigr\rangle_{S_{\rm eff}}}
 {\Bigl\langle e^{i \bigl[{\rm arg}\det\!J_t(x)-S_I(z_t(x))\bigr]} 
 \Bigr\rangle_{S_{\rm eff}}}. 
\label{eq:LT2}
\end{align}
Here, $J_t(x)\equiv \bigl(\partial z_t^i(x)/\partial x^j\bigr)$ is the Jacobi matrix, 
$S_R\bigl(z_t(x)\bigr)$ ($S_I\bigl(z_t(x)\bigr)$) is the real (imaginary) part of $S\bigl(z_t(x)\bigr)$, 
and $\langle \ast \rangle_{S_{\rm eff}}$ is the expectation value 
taken with respect to 
\begin{align}
 S_{\rm eff}(x;t) \equiv S_R\bigl(z_t(x)\bigr) - \ln\,\bigl|\det\!J_t(x)\bigr|.
\label{eq:Seff}
\end{align}
The Jacobi matrix is obtained 
by solving the combined differential equations with respect to $t$ 
\cite{Alexandru:2015xva}:%
\footnote{
The second equation is obtained by differentiating the first equation 
with respect to $x^j\in\bbR$; 
\begin{align}
 \frac{d(J_t(x))_{ij}}{dt}
 =\frac{d}{dt}\Bigl(\frac{\partial z_t^i(x)}{\partial x^j}\Bigr)
 =\frac{\partial}{\partial x^j}
 \Bigl[\frac{\partial S}{\partial z^i}(z_t(x))\Bigr]^\ast
 =\Bigl[\frac{\partial^2 S}{\partial z^i \partial z^k}(z_t(x))\,
 \frac{\partial z^k_t(x)}{\partial x^j}\Bigr]^\ast
 =\Bigl[H_{ik}(z_t(x))\,(J_t)_{kj}(x) \Bigr]^\ast. 
\nonumber
\end{align}
} 
\begin{align}
 \frac{d z_t^i}{d t} 
 &= \Bigl[\frac{\partial S}{\partial z^i}(z_t)\Bigr]^\ast,
 \quad
 z_0^i = x^i, 
\label{eq:flow1}
\\
\frac{d {(J_t)}_{ij}}{d t}  &= \Bigl[ H_{ik}(z_t)\cdot (J_t)_{kj}\Bigr]^\ast,
 \quad
 (J_0)_{ij} = \delta_{ij}, 
\label{eq:flow2}
\end{align}
where $H_{ij}(z)\equiv \partial^2 S(x)/\partial z^i \partial z^j$ 
is the Hesse matrix for the action $S(z)$.

The key point is that 
the right-hand side of \eqref{eq:LT1} or \eqref{eq:LT2} 
does not depend on the flow time $t$, 
so that one can set $t$ to an arbitrary value 
that is convenient for actual calculation. 
Note that under the flow 
the real part $S_R\bigl(z_t(x)\bigr)$ does not decrease 
while the imaginary part $S_I\bigl(z_t(x)\bigr)$ is kept constant, 
because $(d/dt) S\bigl(z_t(x)\bigr)
 =\bigl|\partial_{z^i}S\bigl(z_t(x)\bigr)\bigr|^2\geq 0$. 
In the original Lefschetz thimble algorithm, 
one takes the limit $t\to\infty$, 
in which 
$\Sigma_t$ approaches a union of connected components (Lefschetz thimbles),  
and the action has a constant imaginary part on each thimble.%
\footnote{
Generically there is a single critical point $z_\sigma$ 
on each connected component $\mathcal{J}_\sigma$, 
and $\mathcal{J}_\sigma$ is obtained as the set of orbits 
flowing out of $z_\sigma$. 
The complementary submanifold to $\mathcal{J}_\sigma$ in $\bbC^N$ 
consists of orbits that flow into $z_\sigma$, 
and will be denoted by $\mathcal{K}_\sigma$. 
The integrations in \eqref{eq:LT2} are dominated by points 
near the intersection of $K_\sigma$ and $\bbR^N$.
} 
In a generic situation 
the phase change coming from $J_t(x)$ is sufficiently mild, 
so that the Monte Carlo calculation for the expression \eqref{eq:LT2} 
is free from sign problems. 
However, two different thimbles are also disconnected 
in the sense of Monte Carlo sampling 
because $S_R$ increases indefinitely near the boundary of each thimble. 
This multimodality of distribution makes the Monte Carlo calculation impractical, 
especially when contributions from more than one thimble 
are relevant to estimating expectation values.

A very interesting proposal made in Refs. \cite{Alexandru:2015sua}
is to use a finite amount $t$, 
which is chosen to be large enough to avoid the sign problem 
but also not too large in order to enable 
exploration in the configuration space. 
However, one does not know a priori 
whether the adopted value of $t$ is actually free from 
the two obstacles (the sign and multimodal problems) simultaneously. 
We will show that one can solve both simultaneously 
if we implement the parallel tempering method in their algorithm 
with the flow time as a tempering parameter.

\subsection{Implementation of parallel tempering}
\label{subsec:PTMT}

The basic idea of the parallel tempering algorithm 
\cite{Swendsen1986,Geyer1991,Earl2005} is the following. 
Suppose that we want to estimate expectation values 
with action $S(x;\lambda)$, 
where $x\in\bbR^N$ is a dynamical variable and $\lambda$ 
is the parameter (such as the temperature) 
that we want to use for a Monte Carlo calculation. 
The point is that 
even when the distribution is multimodal for the original $\lambda$ 
(e.g., when $\lambda$ represents a very low temperature), 
the multimodality can be made mild 
if one takes another value $\tilde\lambda$ 
(e.g., $\tilde\lambda$ corresponding to a very high temperature). 
So, if the configuration space is enlarged 
such that the parameter can change gradually 
between $\lambda$ and $\tilde\lambda$, 
two separate configurations for the original $\lambda$ 
will be connected 
by passing through configurations at parameters near $\tilde\lambda$. 
The parallel tempering algorithm enables 
the move of configurations among different $\lambda$ 
by enlarging the configuration space from $\bbR^N=\{x\}$ 
to the set of $A+1$ replicas, $(\bbR^N)^{A+1}=\{(x_0,x_1,\ldots,x_A)\}$. 
We there assign $\lambda_\alpha$ to replica $\alpha$ ($\alpha=0,1,\ldots,A$), 
such that $\lambda_0=\tilde\lambda$ and $\lambda_A=\lambda$ 
and that $\lambda_\alpha$ and $\lambda_{\alpha+1}$ 
are sufficiently close to each other.%
\footnote{
 The computational cost required for the parallel tempering method 
 can be roughly estimated to be proportional to $(A+1)/X$, 
 where $X$ is the minimum of the acceptance rates 
 for all pairs of adjacent replica (see Step~3 below). 
 In this paper, 
 we will set the parameters $A$ and ${\lambda_\alpha}$ 
 such that the minimum acceptance rate is well above $50\%$ 
 (see Fig.~\ref{fig:MT_connectivity}).
} 
We set up an irreducible, aperiodic Markov chain for the enlarged configuration space 
such that the probability distribution for $(x_0,x_1,\ldots,x_A)$ 
eventually approaches the equilibrium distribution 
proportional to 
\begin{align}
 \prod_\alpha e^{-S(x_\alpha;\lambda_\alpha)}.
\end{align}
We finally take sample averages only with respect to 
a sample taken from $\alpha=A$. 
The simplest algorithm to realize this idea%
\footnote{ 
Of course, there can be many variations on this algorithm. 
} 
is to swap two configurations of two adjacent replicas 
$\alpha$ and $\alpha+1$, 
(i.e., 
to update the configuration $(x_\alpha=x,\,x_{\alpha+1}=x')$ 
to $(x_\alpha=x',\,x_{\alpha+1}=x)$)
with the probability 
\begin{align}
 w_\alpha(x,x')={\rm min}\Bigl(1,\,
 \frac{e^{-S(x';\lambda_\alpha)-S(x;\lambda_{\alpha+1})}}
 {e^{-S(x;\lambda_\alpha)-S(x',\lambda_{\alpha+1})}}\Bigr), 
\end{align}
which obviously satisfies the detailed balance condition 
\begin{align}
 w_\alpha(x,x')\,e^{-S(x;\lambda_\alpha)-S(x',\lambda_{\alpha+1})}
 = w_\alpha(x',x)\,e^{-S(x';\lambda_\alpha)-S(x,\lambda_{\alpha+1})}.
\end{align}

Our proposal is to take the flow time $t$ as such a tempering parameter. 
The basic algorithm is then as follows.%
\footnote{
To make discussions simple, we only take the flow time as a tempering parameter. 
The algorithm can be readily extended such that other parameters 
are included as extra tempering parameters. 
} 

\begin{itemize}

\item{\underline{Step 0}.}
Fix the maximum flow time $T$, 
which should be sufficiently large 
such that the sign problem disappears there, 
and pick up flow times $\{t_\alpha\}$ from the interval $[0,T]$ 
with $t_0=0 < t_1 < \cdots < t_A = T$. 
The values of $A$ and $t_\alpha$ are determined manually or adaptively
to optimize the acceptance rate in Step 3 below. 

\item{\underline{Step 1}.}
Choose an initial value $x_\alpha \in \bbR^N$ for each replica $\alpha$, 
and numerically solve the differential equations 
\eqref{eq:flow1} and \eqref{eq:flow2} 
to obtain the triplet 
$(x_\alpha,z_{t_\alpha},J_{t_\alpha})$.

\item{\underline{Step 2}.}
For each $\alpha$, 
construct a Metropolis process to update the value of $x$. 
Explicitly, we take a value $x'_\alpha$ from $x_\alpha$ 
using a symmetric proposal distribution, 
and recalculate the triplet $(x'_\alpha,z'_{t_\alpha},J'_{t_\alpha})$ 
using $x'_\alpha$ as the initial value. 
We then update $x_\alpha$ to $x'_\alpha$ 
with the probability ${\rm min}(1,e^{-\Delta S_{{\rm eff},\alpha}})$, 
where
\begin{align}
 \Delta S_{{\rm eff},\alpha}
 &\equiv S_{\rm eff}(x'_\alpha,t_\alpha)-S_{\rm eff}(x_\alpha,t_\alpha)
\nonumber\\
 &=S_R(z'_{t_\alpha})-\ln\,\bigl|\det\!J'_{t_\alpha}\bigr|
 -S_R(z_{t_\alpha})+\ln\,\bigl|\det\!J_{t_\alpha}\bigr|
\end{align}
(recall that $S_{\rm eff}(x;t) 
= S_R\bigl(z_t(x)\bigr) - \ln\,\bigl|\det\!J_t(x)\bigr|$, 
Eq.~\eqref{eq:Seff}).
We repeat the process sufficiently many times 
such that local equilibrium is realized for each $\alpha$. 

\item{\underline{Step 3}.}
Starting from $\alpha=0$ through $\alpha=A-1$, 
swap the values of $x$ between two adjacent replicas $\alpha$ and $\alpha+1$ 
with the probability 
\begin{align}
 w_\alpha(x_\alpha,x_{\alpha+1})
 \equiv {\rm min}\Bigl(1,\,
 \frac{e^{-S_{\rm eff}(x_{\alpha+1};\,t_\alpha)
 -S_{\rm eff}(x_{\alpha};\,t_{\alpha+1})}}
 {e^{-S_{\rm eff}(x_{\alpha};\,t_\alpha)
 -S_{\rm eff}(x_{\alpha+1};\,t_{\alpha+1)}}}\Bigr).
\end{align}

\item{\underline{Step 4}.}
After repeating Steps 2 and 3 sufficiently many times, 
get a triplet $(x_A,z_{t_A=T},J_{t_A=T})$ from $\alpha=A$ 
as an element of a sample. 

\item{\underline{Step 5}.}
After repeating Steps 2 to 4, 
we obtain a sequence of triplets 
$\{(x_A^{(a)},z_{t_A=T}^{(a)},J_{t_A=T}^{(a)})\}\}$, 
which we use to estimate the expectation value:
\begin{align}
 \langle \mathcal{O}(x) \rangle \thickapprox
 \frac{\sum_a
 e^{i \bigl[{\rm arg}\det\!J_{T}^{(a)}-S_I(z_{T}^{(a)})\bigr]} 
 \mathcal{O}\bigl(z_{T}^{(a)}\bigr)
  }
 {\sum_a
 e^{i \bigl[{\rm arg}\det\!J_{T}^{(a)}-S_I(z_{T}^{(a)})\bigr]} 
 }. 
\label{eq:sum}
\end{align}

\end{itemize}

Note that the action with $t_0=0$ is the original action 
for which the sign problem exists. 
However, as can be seen from \eqref{eq:sum}, 
the sample average is taken 
only with respect to the action with $t_A=T$, 
so that we need not worry about the original sign problem, 
although we include configurations near $t_A=0$. 
Also, $t_A=T$ can be taken to be sufficiently large 
such that the sign problem disappears 
if we complement intermediate flow times sufficiently (with larger $A$). 
Thus, this simple algorithm solves the two obstacles simultaneously: 
the original sign problem at $t_0=0$ 
is resolved at $t_A=T$ 
while the multimodal problem at $t_A=T$ 
is resolved by passing through configurations near $t_0=0$.%
\footnote{
We have implicitly assumed 
that the action at $t_0=0$ does not cause a multimodality 
in the configuration space. 
If this is not the case, 
we further introduce other parameters 
(such as the coefficient of the action) as extra tempering parameters 
or prepare flow times $\{t_\alpha\}$ such that $t_0 < 0$.
} 

\section{Example}
\label{sec:Example}

In this section 
we investigate the $(0+1)$-dimensional massive Thirring model 
at finite density \cite{Pawlowski:2013pje,Pawlowski:2014ada,Fujii:2015bua} 
to exemplify that the algorithm given in the previous section does work. 

The $(0+1)$-dimensional massive Thirring model 
is defined from the standard $(1+1)$-dimensional massive 
Thirring model by dimensional reduction. 
With an auxiliary field $\phi(\tau)$, 
the continuum representation of the grand partition function 
$Z = {\rm tr\,} e^{-\beta(H-\mu Q)}$
is given by the path integral
\begin{align}
 Z = \int_{\rm PBC}[d\phi(\tau)] 
 \int_{\rm ABC}[d\bar\psi(\tau) d\psi(\tau)]\,
 e^{-S[\phi,\bar\psi,\psi]},
\end{align}
where the Euclidean action takes the form
\begin{align}
 S[\phi,\bar\psi,\psi]=\int_0^\beta d\tau\,\Bigl[
 -\bar\psi\bigl( \gamma^0(\partial_0+i\phi+\mu) + m\bigr)\psi
 +\frac{1}{2g^2}\,\phi^2\Bigr],
 \quad
 \biggl[ \gamma^0=\begin{pmatrix} 0&1\\ 1&0\end{pmatrix} \biggr],
\end{align}
and the $\phi$ integral (the $\psi$, $\bar\psi$ integral) 
obeys the periodic (antiperiodic) boundary condition. 
In realizing the model on the lattice, 
we discretize the Euclidean time as $\tau=n a$ $(n=1,\ldots,N)$ 
with $\beta=N a$ ($N$: even), 
and follow the prescription of Ref. \cite{Pawlowski:2014ada}, 
where $\phi(\tau)=\phi_n$ is treated as a $U(1)$ gauge potential 
and is combined with the chemical potential $\mu$ 
to become a link variable of a complexified gauge group, 
$e^{(i\phi(\tau)+\mu)a} = e^{i\phi_n a}e^{\mu a}\equiv U_n\,e^{\mu a}$, 
as proposed in Ref. \cite{Hasenfratz:1983ba}. 
Then, by using a staggered fermion formulation, 
the grand partition function is given by 
\begin{align}
 Z = \int (d U)\,e^{-S(U,\bar\chi,\chi)},
\end{align}
where $(dU)=\prod_n(dU_n)\equiv\prod_n [d(\phi_n a)/2\pi]$ 
(with $\phi_n a\in (-\pi,\pi]$) 
and 
\begin{align}
 S(U,\bar\chi,\chi) = -\sum_{n,\ell=1}^N \bar\chi_n\, D_{n\ell}(U)\,\chi_\ell
 +\frac{1}{2g^2 a}\,\sum_{n=1}^N \Bigl[1-\frac{1}{2}(U_n+U_n^{-1})\Bigr]
\end{align}
with 
$D_{n\ell}(U) \equiv (1/2)\bigl(U_n e^{\mu a} \delta_{n+1,\ell}
 -U_{n-1}^{-1} e^{-\mu a} \delta_{n-1,\ell}
 -U_N e^{\mu a} \delta_{n,N}\delta_{\ell,1}
 +U_N^{-1} e^{-\mu a} \delta_{n,1}\delta_{\ell,N}\bigr)
 + ma\,\delta_{n\ell}$.
We henceforth set $a=1$ and treat $m$, $\mu$, 
and $g^2$ as dimensionless parameters.

After carrying out the fermion integration, we obtain
\begin{align}
 Z &= \int (d U)\,\det D(U)\,
 e^{-(1/2g^2)\sum_n[1-(1/2)(U_n+U_n^{-1})]},
\end{align}
which can be calculated analytically \cite{Pawlowski:2014ada} as 
\begin{align}
 Z = \frac{e^{-N\alpha}}{2^{N-1}}\bigl[
 \cosh(N\mu)\,I_1^N(\alpha) + \rho_+\,I_0^N(\alpha)\bigr],
\end{align}
where $\alpha\equiv 1/(2g^2)$, 
$2\rho_\pm\equiv (\sqrt{m^2+1}+m)^N \pm (\sqrt{m^2+1}-m)^N$, 
and $I_n(\alpha)$ is the modified Bessel function of the first kind 
of order $n$.
Using this expression, 
we obtain the analytic form of the chiral condensate 
as
\begin{align}
 \langle \bar\chi \chi \rangle
 &=\frac{1}{N}\frac{\partial \ln Z}{\partial m}
 =\frac{\rho_-\, I_0^N(\alpha)}{\sqrt{m^2+1}\,
   \bigl[\cosh(N\mu)\, I_1^N(\alpha) + \rho_+\, I_0^N(\alpha)\bigr]}.
\label{eq:chiralcond_anal}
\end{align}

We now numerically evaluate the chiral condensate 
by using the complex action 
\begin{align}
 S(U) \equiv \frac{1}{2g^2}\,\sum_n\Bigl[1-\frac{1}{2}\,(U_n+U_n^{-1})\Bigr]
 -\ln\det D(U)
\label{eq:chichi_anal}
\end{align}
as
\begin{align}
 \langle \bar\chi \chi \rangle
 = \Bigl\langle \frac{1}{N}\,{\rm tr\,}
 \Bigl(D^{-1}(U)\frac{\partial D(U)}{\partial m}\Bigr)
 \Bigr\rangle
 = \Bigl\langle \frac{1}{N}\,{\rm tr\,} D^{-1}(U)
 \Bigr\rangle,
\end{align}
where
\begin{align}
 \langle \mathcal{O}(U) \rangle 
 \equiv \frac{\int (dU )\, e^{-S(U)}\,\mathcal{O}(U)}
 {\int (dU)\, e^{-S(U)}}.
\end{align}
It is easy to check that $[\det D(U;\mu)]^\ast = \det D(U;-\mu)$, 
so that the second term of \eqref{eq:chichi_anal} is complex-valued  
for real $\mu$, 
and the sign problem should arise when $N$ is large.

\begin{figure}[htbp]
\begin{center}
\includegraphics[width=8cm]{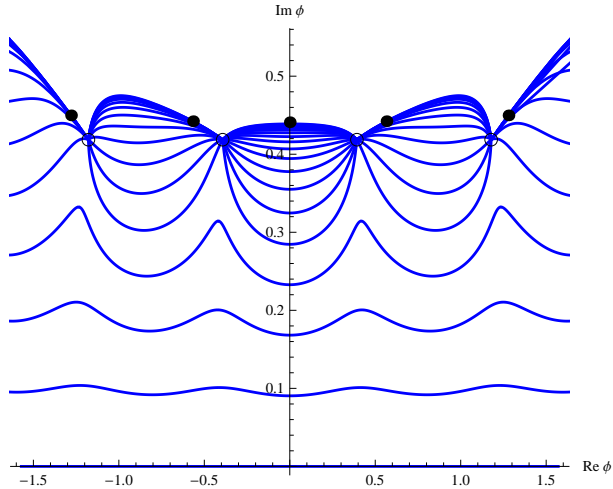}
\begin{quote}
\caption{$\phi(t)=(1/N)\sum_n\phi_n(t)$  
at flow times $t=0,\,0.1,\,0.2,\ldots,1.9,\,2.0$ from bottom to top. 
The parameters are set to $N=8$, $m=1$, $\mu=1.3$, $g^2=1/2$.
The full circles are the critical points of $S$, 
and the empty circles are the log singularities of $S$. 
}
\label{fig:MT_flow}
\end{quote}
\end{center}
\vspace{-6ex}
\end{figure}

\begin{figure}[htbp]
\begin{center}
\includegraphics[width=12cm]{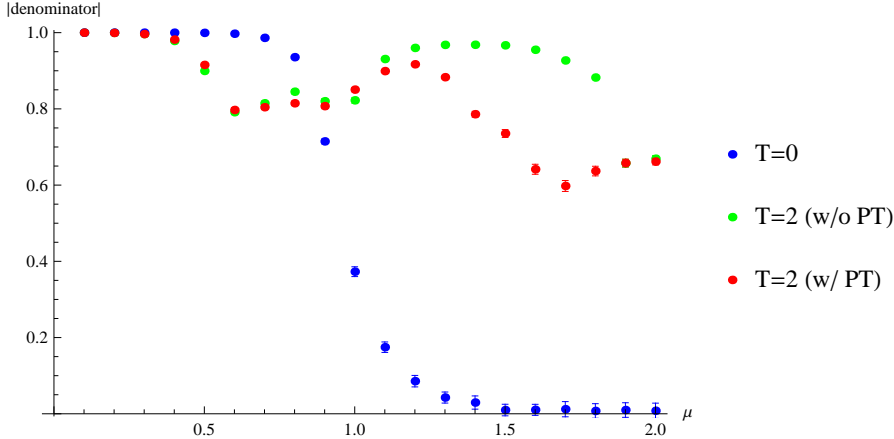}
\begin{quote}
\caption{The absolute value of the denominator in \eqref{eq:sum} 
(divided by the sample size)
as a function of $\mu$ 
with the other parameters set to $N=8$, $m=1$, $g^2=1/2$.
The blue points are the result for the flow time $T=0$ 
and show that the sign problem actually exists for $\mu \gtrsim 1.0$. 
The green (red) points are the result 
for the flow time $T=2$ without (with) the parallel tempering (PT) implemented, 
and show that the sign problem disappears at $T=2$. 
}
\label{fig:MT_denom}
\end{quote}
\end{center}
\vspace{-6ex}
\end{figure}

We follow the steps given in Sect.~\ref{subsec:PTMT} 
with the complexification of the variables $\phi_n\in (-\pi,\pi]$. 
We first set the largest flow time to $T=2$, 
and prepare flow times $\{t_\alpha\}$ $(\alpha=0,1,2,\ldots,20)$ 
with equal separations as 
$t_0=0,\,t_1=0.1,\,t_2=0.2,\,\ldots,t_{20}=2.0$.%
\footnote{
 The fermion determinant at flow time $t=0$ is given by
 $\det D=(1/2^N)\,[\zeta + \zeta^{-1}+(\sqrt{2}+1)^N+(\sqrt{2}-1)^N)]$
 where $\zeta\equiv e^{i\sum_n\phi_n+N\mu}$.
 Note that  
 $\det D$ vanishes at $\sum_n \phi_n=\pi$ $(\mbox{mod}~2\pi)$
 when $\mu=\log(\sqrt{2}+1)\sim 0.881$,
 which gives a multimodal distribution even at $t=0$.
 However, this can be handled without introducing another tempering parameter,
 because each mode has a rather wide distribution 
 (with not-too-small values near the boundary) 
 and thus the whole configuration space can be easily explored
 by setting the interval of proposal distribution
 to be a large value, as in this paper. 
} 
Figure~\ref{fig:MT_flow} shows 
$\phi(t)\equiv(1/N)\sum_n\phi_n(t)$  
at flow times $t=t_\alpha$ 
with the $\phi_n(0)$ set to the same value.
As Step 1 of our algorithm, 
we make a cold start ($\phi_n=0$) for every replica $\alpha$, 
and numerically solve the differential equations 
\eqref{eq:flow1} and \eqref{eq:flow2} 
with the adaptive 4th-order Runge--Kutta method 
to obtain the triplet $(x_\alpha,z_{t_\alpha},J_{t_\alpha})$. 
We repeat the Metropolis process twenty times in Step 2,%
\footnote{ 
As a proposal distribution 
we use the uniform distribution within the interval $[-\epsilon,\epsilon]$, 
where $\epsilon$ is chosen randomly from 
$\{1,\,10^{-1},\,10^{-2},\ldots,\,10^{-[2t_\alpha+1]}\}$ 
($[k]$ is the floor of $k$). 
} 
which is followed by a single sequence of swapping of Step 3. 
We then repeat Steps 2 and 3 ten times (as Step 4). 
With the first 5 data points discarded as initial sweeps, 
we estimate correlation functions with $10^4$ data points. 
Figure~\ref{fig:MT_denom} shows the absolute value 
of the denominator in \eqref{eq:sum} (divided by the sample size)
as a function of $\mu$ 
with the other parameters set to $N=8$, $m=1$, $g^2=1/2$.  
The blue points are the result for the flow time $T=0$. 
They correspond to the usual reweighting calculus, 
and show that the sign problem actually exists for $\mu\gtrsim 1.0$. 
The green (red) points are the result 
for the flow time $T=2$ without (with) the parallel tempering implemented. 
The results show that the sign problem disappears at $T=2$.

\begin{figure}[htbp]
\begin{center}
\includegraphics[width=13cm]{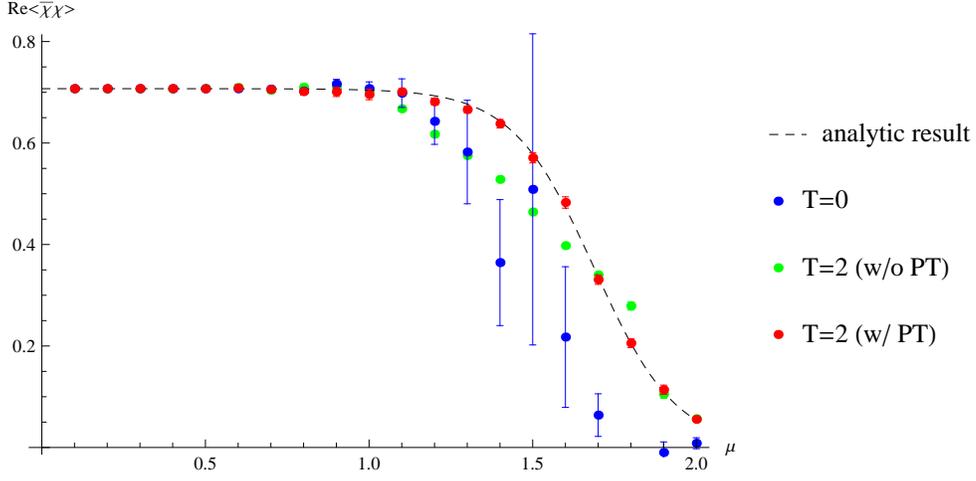}
\begin{quote}
\caption{Chiral condensate $\langle\bar\chi \chi\rangle$ 
as a function of $\mu$  
with the other parameters set to $N=8$, $m=1$, $g^2=1/2$.
The dotted line represents the analytic result \eqref{eq:chiralcond_anal}. 
The blue (green) points are the result for the flow time $T=0$ ($T=2$) 
without the parallel tempering implemented. 
The red points are the result for the flow time $T=2$ 
with the parallel tempering implemented. 
}
\label{fig:MT_chiralcond}
\end{quote}
\end{center}
\vspace{-6ex}
\end{figure}

\begin{figure}[htbp]
\begin{center}
\includegraphics[width=6cm]{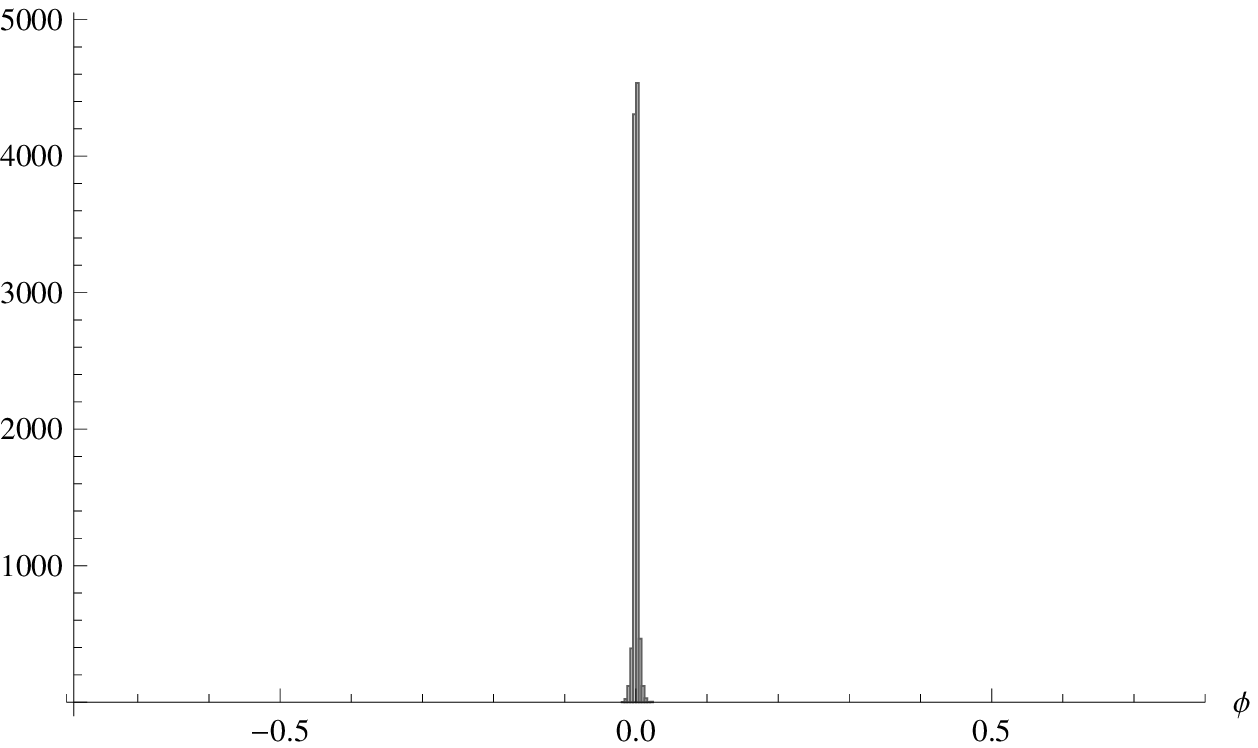}
\hspace{5mm}
\includegraphics[width=6cm]{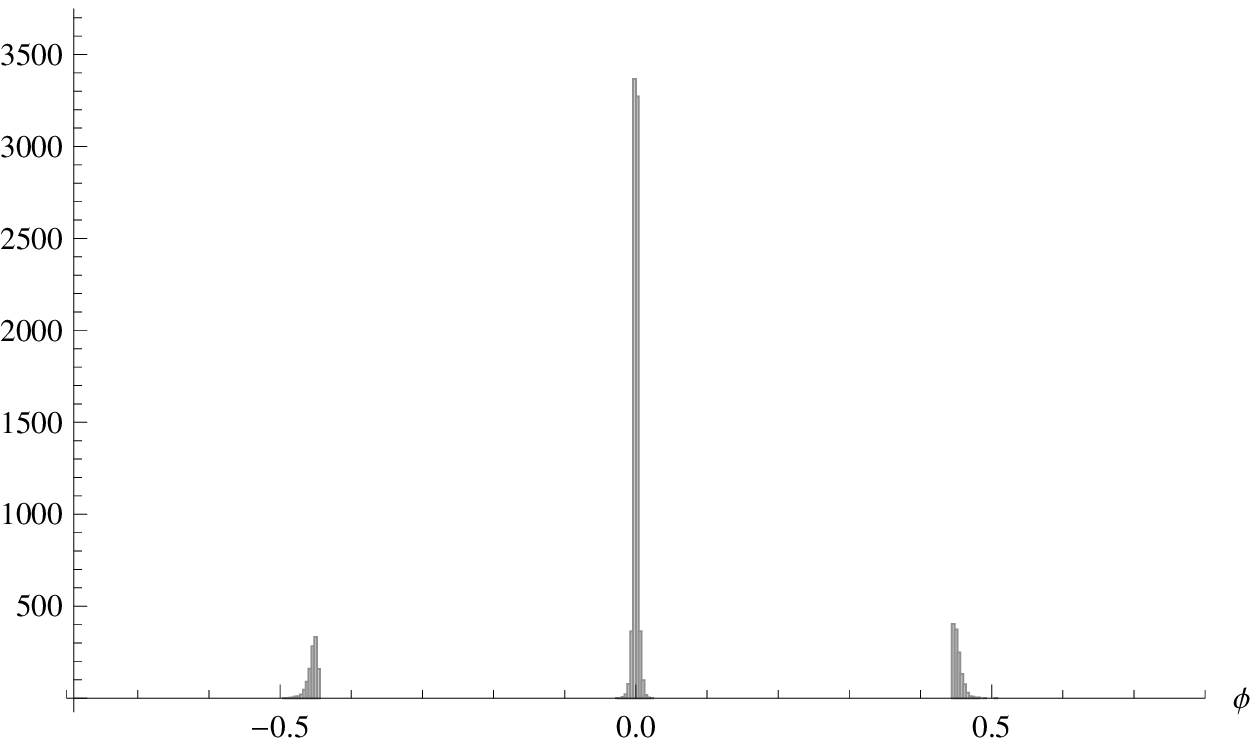}
\begin{quote}
\caption{Histograms of $\phi= (1/N)\sum_n \phi_n$ 
at the flow time $T=2$ 
without/with the parallel tempering implemented (left/right). 
The parameters are set to $N=8$, $m=1$, $\mu=1.3$, $g^2=1/2$. 
}
\label{fig:MT_histogram}
\end{quote}
\end{center}
\vspace{-6ex}
\end{figure}

\begin{figure}[htbp]
\begin{center}
\includegraphics[width=6cm]{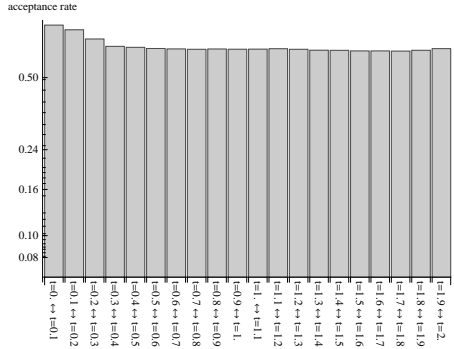}
\begin{quote}
\caption{Average acceptance rate 
for the swap between replicas $\alpha$ and $\alpha+1$ 
in Step 3. 
The parameters are set to $N=8$, $m=1$, $\mu=1.3$, $g^2=1/2$. 
}
\label{fig:MT_connectivity}
\end{quote}
\end{center}
\vspace{-6ex}
\end{figure}

Figure~\ref{fig:MT_chiralcond} 
shows the chiral condensate $\langle\bar\chi \chi\rangle$ 
as a function of $\mu$. 
The other parameters are again set to $N=8$, $m=1$, $g^2=1/2$. 
The dotted line represents the analytic result \eqref{eq:chiralcond_anal}. 
The blue points are the result for the flow time $T=0$ 
and exhibit large statistical errors, reflecting the sign problem. 
The green points are the result for the flow time $T=2$ 
without the parallel tempering implemented. 
They have small statistical errors, 
but exhibit statistically significant discrepancies from the analytic result. 
This should be attributed, 
as discussed in detail in Ref. \cite{Alexandru:2015sua}, 
to the fact 
that the dominant contributions come only from a single thimble 
for such large $T$.  
The red points are the result for the flow time $T=2$ 
now with the parallel tempering implemented. 
They show a good agreement with the analytic result, 
which implies that contributions from various thimbles 
are correctly taken into account 
through the parallel tempering. 
This can be confirmed by Figs.~\ref{fig:MT_histogram} 
and \ref{fig:MT_connectivity}. 
Figure~\ref{fig:MT_histogram} exhibits the histogram 
of $\phi= (1/N)\sum_n \phi_n$ 
at $T=2$. 
We see that the configurations are concentrated on a single thimble 
if the parallel tempering is not implemented (left), 
while they are spread out on various thimbles 
if the parallel tempering is implemented (right). 
Figure~\ref{fig:MT_connectivity} shows the average acceptance rates 
for the swaps between replicas $\alpha$ and $\alpha+1$ 
in Step 3. 
We see that the swapping is carried out very well 
because the average acceptance rate is 
more than $50\%$ for all pairs $(\alpha,\alpha+1)$.

\section{Conclusion and outlook}
\label{sec:Conclusion}

In this paper, as a versatile tool 
for Monte Carlo calculations of models with complex actions,  
we have proposed a Lefschetz thimble algorithm 
where the flow time is used as an auxiliary variable 
for the highly multimodal distribution. 
In particular, we implemented the parallel tempering method 
by taking the flow time $t$ as a tempering parameter. 
There, we prepare flow times $\{t_\alpha\}$ 
such that $t_0=0$ and $t_A=T$. 
The largest flow time $T$ can be taken to be sufficiently large 
such that the sign problem disappears there. 
Although the algorithm includes configurations at $t_0=0$, 
the original sign problem at $t_0=0$ does not enter the calculation 
because the sample average is taken only with respect to $t_A=T$, 
where the sign problem disappears. 
We have investigated the $(0+1)$-dimensional massive Thirring model 
at finite density to exemplify that the algorithm does work, 
and showed that contributions from multi thimbles 
are correctly taken into account 
even for such a large flow time as $T=2$.

We should investigate to what extent this algorithm is actually versatile. 
One interesting class of models, for which we can readily test our algorithm 
before applying it to QCD at finite density, 
is that of various types of large-$N$ random matrix models with complex actions. 
In fact, if the free energy is calculated 
by an integration over matrices themselves (not over their eigenvalues), 
the classical solutions and the corresponding Lefschetz thimbles 
do not have a useful meaning 
because the quantum corrections are 
of the same order as the leading term. 
It thus provides us with a good test of versatility  
to check whether correct results are obtained for such models 
where the thimble structure or its usefulness is not clear. 
As a related model, 
the numerical study of the triangle--hinge model 
\cite{Fukuma:2015xja,Fukuma:2015haa,Fukuma:2016zea} 
should also be interesting. 
The model is a sort of matrix model 
that generates 3D random volumes 
as a collection of triangles and hinges. 
In order to restrict the resulting configurations 
to tetrahedral decompositions, 
one needs to introduce a special form of interaction \cite{Fukuma:2015xja}, 
which makes the action complex-valued 
(M. Fukuma, S. Sugishita, and N. Umeda, manuscript in preparation). 
A numerical study was made for a simplified model 
(with no restriction to tetrahedral decompositions), 
and the existence of a third-order phase transition is confirmed 
(manuscript in preparation). 
It is thus interesting to see whether the phase transition still exists 
when the restriction is imposed.

Besides the Lefschetz thimble algorithm, 
the complex Langevin algorithm \cite{Parisi:1984cs,Aarts:2013uxa}
is also under intense study 
as a promising method to solve the sign problem. 
Recently, a very interesting proposal was made 
by Bloch \cite{Bloch:2017ods} (see also Ref. \cite{Bloch:2016jwt})
to evaluate correlation functions 
by reweighting complex Langevin trajectories 
using such parameters that satisfy known validity conditions 
\cite{Aarts:2011ax,Nagata:2016vkn}
to be free from wrong convergence problems 
\cite{Aarts:2011ax,Hayata:2015lzj,Nagata:2016vkn,Abe:2016hpd,Salcedo:2016kyy}.  
It should be interesting to compare the extent of versatility 
between the reweighted complex Langevin algorithm 
and our parallel tempering algorithm with the flow time 
as a tempering parameter. 

A study along these lines is now in progress 
and will be reported elsewhere.

\section*{Acknowledgements}
The authors thank Hikaru Kawai, Jun Nishimura, 
Sotaro Sugishita, and Asato Tsuchiya
for useful discussions. 
This work was partially supported by MEXT (Grant No.\,16K05321).



\baselineskip=0.9\normalbaselineskip



\begin{thebibliography}{99}
\setlength{\itemsep}{-2pt}

\bibitem{Aarts:2015tyj} 
  G.~Aarts,
  ``Introductory lectures on lattice QCD at nonzero baryon number,''
  J.\ Phys.\ Conf.\ Ser.\  {\bf 706}, no. 2, 022004 (2016)
  [arXiv:1512.05145 [hep-lat]].

\bibitem{Cristoforetti:2012su} 
  M.~Cristoforetti {\it et al.} [AuroraScience Collaboration],
  ``New approach to the sign problem in quantum field theories: High density QCD on a Lefschetz thimble,''
  Phys.\ Rev.\ D {\bf 86}, 074506 (2012)
  [arXiv:1205.3996 [hep-lat]].

\bibitem{Cristoforetti:2013wha} 
  M.~Cristoforetti, F.~Di Renzo, A.~Mukherjee and L.~Scorzato,
  ``Monte Carlo simulations on the Lefschetz thimble: Taming the sign problem,''
  Phys.\ Rev.\ D {\bf 88}, no. 5, 051501 (2013)
  [arXiv:1303.7204 [hep-lat]].
 
\bibitem{Mukherjee:2013aga} 
  A.~Mukherjee, M.~Cristoforetti and L.~Scorzato,
  ``Metropolis Monte Carlo integration on the Lefschetz thimble: Application to a one-plaquette model,''
  Phys.\ Rev.\ D {\bf 88}, no. 5, 051502 (2013)
  [arXiv:1308.0233 [physics.comp-ph]].
  
\bibitem{Fujii:2013sra} 
  H.~Fujii, D.~Honda, M.~Kato, Y.~Kikukawa, S.~Komatsu and T.~Sano,
  ``Hybrid Monte Carlo on Lefschetz thimbles - A study of the residual sign problem,''
  JHEP {\bf 1310}, 147 (2013)
  [arXiv:1309.4371 [hep-lat]].

\bibitem{Cristoforetti:2014gsa} 
  M.~Cristoforetti, F.~Di Renzo, G.~Eruzzi, A.~Mukherjee, C.~Schmidt, L.~Scorzato and C.~Torrero,
  ``An efficient method to compute the residual phase on a Lefschetz thimble,''
  Phys.\ Rev.\ D {\bf 89}, no. 11, 114505 (2014)
  [arXiv:1403.5637 [hep-lat]].

\bibitem{Alexandru:2015xva} 
  A.~Alexandru, G.~Ba\c sar and P.~Bedaque,
  ``Monte Carlo algorithm for simulating fermions on Lefschetz thimbles,''
  Phys.\ Rev.\ D {\bf 93}, no. 1, 014504 (2016)
  [arXiv:1510.03258 [hep-lat]].

\bibitem{Kanazawa:2014qma} 
  T.~Kanazawa and Y.~Tanizaki,
  ``Structure of Lefschetz thimbles in simple fermionic systems,''
  JHEP {\bf 1503}, 044 (2015)
  [arXiv:1412.2802 [hep-th]].
  
\bibitem{Tanizaki:2015rda} 
  Y.~Tanizaki, Y.~Hidaka and T.~Hayata,
  ``Lefschetz-thimble analysis of the sign problem in one-site fermion model,''
  New J.\ Phys.\  {\bf 18}, no. 3, 033002 (2016)
  [arXiv:1509.07146 [hep-th]].
  
\bibitem{Fujii:2015bua} 
  H.~Fujii, S.~Kamata and Y.~Kikukawa,
  ``Lefschetz thimble structure in one-dimensional lattice Thirring model at finite density,''
  JHEP {\bf 1511}, 078 (2015)
  Erratum: [JHEP {\bf 1602}, 036 (2016)]
  [arXiv:1509.08176 [hep-lat]].

\bibitem{Alexandru:2015sua} 
  A.~Alexandru, G.~Ba\c sar, P.~F.~Bedaque, G.~W.~Ridgway and N.~C.~Warrington,
  ``Sign problem and Monte Carlo calculations beyond Lefschetz thimbles,''
  JHEP {\bf 1605}, 053 (2016)
  [arXiv:1512.08764 [hep-lat]].

\bibitem{Alexandru:2016ejd} 
  A.~Alexandru, G.~Ba\c sar, P.~F.~Bedaque, G.~W.~Ridgway and N.~C.~Warrington,
  ``Monte Carlo calculations of the finite density Thirring model,''
  Phys.\ Rev.\ D {\bf 95}, no. 1, 014502 (2017)
  [arXiv:1609.01730 [hep-lat]].

\bibitem{Swendsen1986}
  R.~H.~Swendsen and J.-S.~Wang,
  ``Replica Monte Carlo Simulation of Spin-Glasses,''
  Phys.\ Rev.\ Lett.\ {\bf 57} 2607 (1986). 
  
\bibitem{Geyer1991}
  C.~J.~Geyer, 
  ``Markov Chain Monte Carlo Maximum Likelihood,''
  in Computing Science and Statistics: Proceedings of the 23rd Symposium
  on the Interface, Interface Foundation, Fairfax Station, VA, p.~156 (1991).

\bibitem{Earl2005}
  D.~J.~Earl and M.~W.~Deem, 
  ``Parallel tempering: Theory, applications, and new perspectives,''
  Phys.\ Chem.\ Chem.\ Phys.\ {\bf 7}, 3910 (2005).

\bibitem{Pawlowski:2013pje} 
  J.~M.~Pawlowski and C.~Zielinski,
  ``Thirring model at finite density in 0+1 dimensions with stochastic quantization: Crosscheck with an exact solution,''
  Phys.\ Rev.\ D {\bf 87}, no. 9, 094503 (2013)
  [arXiv:1302.1622 [hep-lat]].
  
\bibitem{Pawlowski:2014ada} 
  J.~M.~Pawlowski, I.~O.~Stamatescu and C.~Zielinski,
  ``Simple QED- and QCD-like Models at Finite Density,''
  Phys.\ Rev.\ D {\bf 92}, no. 1, 014508 (2015)
  [arXiv:1402.6042 [hep-lat]].
  
\bibitem{Hasenfratz:1983ba} 
  P.~Hasenfratz and F.~Karsch,
  ``Chemical Potential on the Lattice,''
  Phys.\ Lett.\  {\bf 125B}, 308 (1983).
  
\bibitem{Fukuma:2015xja} 
  M.~Fukuma, S.~Sugishita and N.~Umeda,
  ``Random volumes from matrices,''
  JHEP {\bf 1507}, 088 (2015)
  [arXiv:1503.08812 [hep-th]].
  
\bibitem{Fukuma:2015haa} 
  M.~Fukuma, S.~Sugishita and N.~Umeda,
  ``Matter fields in triangle-hinge models,''
  PTEP {\bf 2016}, no. 5, 053B04 (2016)
  [arXiv:1504.03532 [hep-th]].

\bibitem{Fukuma:2016zea} 
  M.~Fukuma, S.~Sugishita and N.~Umeda,
  ``Triangle-hinge models for unoriented membranes,''
  PTEP {\bf 2016}, no. 7, 073B01 (2016)
  [arXiv:1603.05199 [hep-th]].
  
\bibitem{Parisi:1984cs} 
  G.~Parisi,
  ``On Complex Probabilities,''
  Phys.\ Lett.\  {\bf 131B}, 393 (1983).

\bibitem{Aarts:2013uxa} 
  G.~Aarts, L.~Bongiovanni, E.~Seiler, D.~Sexty and I.~O.~Stamatescu,
  ``Controlling complex Langevin dynamics at finite density,''
  Eur.\ Phys.\ J.\ A {\bf 49}, 89 (2013)
  [arXiv:1303.6425 [hep-lat]].
  
\bibitem{Bloch:2017ods} 
  J.~Bloch,
  ``Reweighting complex Langevin trajectories,''
  arXiv:1701.00986 [hep-lat].
  
\bibitem{Bloch:2016jwt} 
  J.~Bloch, J.~Glesaaen, O.~Philipsen, J.~Verbaarschot and S.~Zafeiropoulos,
  ``Complex Langevin simulations of a finite density matrix model for QCD,''
  arXiv:1612.04621 [hep-lat].

\bibitem{Aarts:2011ax} 
  G.~Aarts, F.~A.~James, E.~Seiler and I.~O.~Stamatescu,
  ``Complex Langevin: Etiology and Diagnostics of its Main Problem,''
  Eur.\ Phys.\ J.\ C {\bf 71}, 1756 (2011)
  [arXiv:1101.3270 [hep-lat]].

\bibitem{Nagata:2016vkn} 
  K.~Nagata, J.~Nishimura and S.~Shimasaki,
  ``Argument for justification of the complex Langevin method and the condition for correct convergence,''
  Phys.\ Rev.\ D {\bf 94}, no. 11, 114515 (2016)
  [arXiv:1606.07627 [hep-lat]].

\bibitem{Hayata:2015lzj}
  T.~Hayata, Y.~Hidaka and Y.~Tanizaki,
  ``Complex saddle points and the sign problem 
  in complex Langevin simulation,''
  Nucl.\ Phys.\ B {\bf 911} (2016) 94
  [arXiv:1511.02437 [hep-lat]].

\bibitem{Abe:2016hpd}
  Y.~Abe and K.~Fukushima,
  ``Analytic studies of the complex Langevin equation 
  with a Gaussian ansatz and multiple solutions in the unstable region,''
  Phys.\ Rev.\ D {\bf 94} (2016) no.9,  094506
  [arXiv:1607.05436 [hep-lat]].

\bibitem{Salcedo:2016kyy}
  L.~L.~Salcedo,
  ``Does the complex Langevin method give unbiased results?,''
  Phys.\ Rev.\ D {\bf 94} (2016) no.11,  114505
  [arXiv:1611.06390 [hep-lat]].
 
\end{thebibliography}
\end{document}